\documentstyle[11pt]{article}
\begin{document}
\renewcommand{\theequation}{\arabic{equation}}
\bibliographystyle{try}
\baselineskip 12pt
\vspace*{-2cm}
\noindent
UCL-HEP-9302
\vspace{1.8cm}
\begin{center}
{\Large{\bf Production of $\eta$ Mesons Near Threshold}}
\footnotetext{Talk given at the BNL workshop on future directions in particle
and nuclear physics at multi-GeV hadron beam facilities, March 4--6, 1993.}
\\[1ex]
 Colin Wilkin\\
University College London,\\ Gower St., London WC1E 6BT, U.K.\\[2ex]
{\bf Abstract}
\vspace{-0.2cm}
\end{center}
\begin{quote}
It is argued that the strong energy dependence of the $p\, d \rightarrow
\,^{3\!}H\!e\, \eta$  cross section near threshold is due to a final state
interaction between the $\eta$ meson and the $^3$He nucleus. The large
scattering length that this implies is in accord with optical potential
predictions and is evidence for a nearby virtual `$\!$`bound'$\!$' state in
the $\eta^3$He system. This model suggests that sharp structures should also
be seen close to production thresholds on other light nuclei.
\end{quote}

The $p\, d \rightarrow \,^{3\!}H\!e\, \eta$  and $p\, d \rightarrow
\,^{3\!}H\!e\, \pi^{0}$ cross sections show striking but very different energy
dependences for proton energies within a few MeV of their respective production
thresholds. To illustrate this, define average squared amplitudes through
\vspace{-0.2cm}
\begin{equation}
|f_{\eta (\pi)}|^{2} = \frac{p_{p}}{p_{\eta (\pi)}}\,\left(\frac{d\sigma}
{d\Omega}\right)_{\!\rm cm}\: ,
\end{equation}
where the phase space ratio of outgoing to incident centre-of-mass momenta has
been factored out.

If the centre-of-mass angular distributions are parameterised as
\begin{equation}
|f_{\eta(\pi)}|^2 \propto 1 + \alpha_{\eta(\pi)}\,
cos\,\theta_{p\eta(\pi)}\: ,
\end{equation}
then $|\alpha_{\eta}|\leq 0.05$ for
$p_{\eta}\leq 0.38$~fm$^{-1}$ \cite{Garcon}. In contrast
$\alpha_{\pi} = 0.73\pm0.02$ already by $p_{\pi}=0.14$~fm$^{-1}$ \cite{Pickar}.

On the other hand $|f_{\eta}|^{2}$ decreases by over a factor of three between
threshold and $p_{\eta}=0.35$~fm$^{-1}$, i.e. $\Delta T_p  = 10$~MeV
\cite{Garcon,Berger}, but the angular average of $|f_{\pi}|^{2}$ has a
rather weak energy dependence near threshold \cite{Pickar,Kerboul}.

The strong angular dependence seen in
$p\, d \rightarrow \,^{3\!}H\!e\, \pi^{0}$ arises from the large $\pi N$
P-wave, associated with the $\Delta$(1232), interfering with an S-wave which
is only significant within a few MeV of threshold. Taking this into account,
both the cross section and deuteron tensor analysing power
are described quantitatively at low energies by
a model involving a spectator nucleon \cite{GW1}.

The most prominent feature of the low energy \mbox{$\eta\, N$} interaction is
an \underline{L=0} resonance, the S$_{11}$ N$^{*}$(1535), and the low energy
$\pi^{-}\, p \rightarrow \eta\, n$  reaction shows only a weak angular
dependence \cite{Binnie}. I want to convince you that this strong S-wave
resonance is also responsible for the rapid energy variation of the
near-threshold $p\, d \rightarrow \,^{3\!}H\!e\, \eta$  cross section.

The spectator-nucleon model, which was successful in $\pi^0$ production,
underestimates significantly the low energy $p\, d \rightarrow \,^{3\!}H\!e\,
\eta$ cross section \cite{LL,GW2}. By including scattering of $\eta$'s and
$\pi$'s on up to \underline{three} nucleons, Laget and Lecolley \cite{LL}
could enhance the cross section but their {\it renormalised} results, shown in
fig.~1, fall off much less steeply than the data \cite{Garcon}.
It is important to note that such a model is perturbative, treating all
interactions only to lowest order. Though this might be justified for low
energy pions, it is dangerous for S-wave $\eta$-nucleon scattering, where the
interaction is so strong.

The sharpness of the energy scale, combined with the isotropy of the angular
distribution, suggests that an S-wave $\eta$--$^3$He final state interaction
(FSI) is responsible for the phenomena rather than the details of the reaction
mechanism or nuclear form factors.

A common approximation \cite{Gold}, in the case of a weak transition to a
channel with a strong FSI, leads to an enhancement of the S-wave amplitude
of the form
\begin{equation}
f_{\eta} = \frac{f_{\eta}^B}{p_{\eta}\, a_{\eta}\,{\rm cot}\delta_{\eta}
- ip_{\eta}\, a_{\eta}}\ \ .
\end{equation}
The influence of the S$_{11}$ resonance is felt through the S-wave phase shift
$\delta_{\eta}$ and scattering length $a_{\eta}$. A reaction model is needed
to estimate the amplitude $f_{\eta}^B$, but it should be slowly varying for
$p_{\eta}R < 1$.

At low energies it is often sufficient to take
\begin{equation}
\label{simple}
f_{\eta} \approx \frac{f_{\eta}^B}{1 - ip_{\eta}\, a_{\eta}}\ \ .
\end{equation}
This corresponds to imposing unitarity with \underline{constant} K-matrix
elements, {\it i.e.} neglecting effective range effects. At present it is
pointless trying to go further.

One cannot unambiguously extract values of the real ($a_R$) and imaginary
($a_I$) parts of the $\eta\,^{3\!}H\!e$ scattering lengths directly from
the present data using eq.~(\ref{simple}). Taking the updated SPES2 points
\cite{Garcon}, the valley of $\chi^2$ lies roughly along
\begin{equation}
a_{R}^{\, 2} + 0.866 a_{I}^{\, 2} + 2.615 a_{I} = 21.69\: ,
\end{equation}
which demonstrates that the modulus of the scattering length has to be bigger
than the nucleus itself! Of course Laget and Lecolley already have
some FSI in their model \cite{LL}, but only a modest amount associated
with $\eta$'s scattering off individual nucleons, leading to
$a_{\eta}\approx (1.2+i0.6)$~fm.

The difficulty in determining {\it both} real and imaginary parts is clear
from the two fits shown in fig.~1 with $(a_R,\ a_I)$ equal to $(3.10,\ 2.52)$
and $(5.0,\ 0.17)$~fm. Though the former gives a better overall agreement with
the data, the place where the predictions really deviate is for
$p_{\eta}\leq 0.05$~fm$^{-1}$ and here it is the hardest to measure due to
intrinsic beam spread and energy loss in the target. The lowest SPES2 point
is averaged over a wide range of energies and the predicted counting rate
$dR/dT$ shown in fig.~2 has a very skewed peak arising from the product of
rapidly varying acceptance and cross section. A different approach will be
needed to do better than this careful SPES2 measurement \cite{Garcon} and the
one that springs to mind is the Celsius storage ring with its cooled proton
beam and gas jet target \cite{Bo}.

Is such a large scattering length consistent with our (limited) knowledge
of $\eta$-nucleus dynamics? In impulse approximation the $\eta\,^{3}$He
scattering length is about four times that of $\eta N$ but there are major
corrections to this simple ansatz due to the strength of the interaction. A
more reliable starting point is to consider the lowest order $\eta\,^{3}$He
optical potential for which
\begin{equation}
\label{opt}
2m_{\eta\, N}^{R}\: V_{\rm opt}(r) = -4\pi\, A\,\rho (r)\,
a\, (\eta\, N)\: ,
\end{equation}
where $m_{\eta\, N}^R$ is the $\eta$-nucleon reduced mass and A (= 3) the mass
number.

Bhalerao and Liu \cite{Liu} analysed the $\pi N$ and $\eta N$ coupled channels
near the $\eta$ threshold within an isobar model and from the $\pi N$ phase
shifts extracted a value for the $\eta$-nucleon scattering length of
$a\, (\eta N) = (0.27 + i0.22)$~fm.. However it is more consistent to apply
eq.~(\ref{simple}) directly to $\pi^{-}\, p \rightarrow \eta\, n$ data.

Using detailed balance and the optical theorem, a lower bound on the imaginary
part of $a\, (\eta N)$ is provided by the threshold $\pi^{-}\, p \rightarrow
\eta\, n$  cross section:
\begin{equation}
{\rm Im}[a\, (\eta N)]\ \geq\ \frac{3}{8\pi}\: \frac{p_{\pi}^{2}}{p_{\eta}}\:
\sigma_{\rm tot}(\pi^{-}\, p \rightarrow \eta\, n)\: .
\end{equation}
The data of ref.~\cite{Binnie} require
${\rm Im}[a\, (\eta N)] \geq (0.28 \pm 0.04)$~fm which, though
larger than that extracted in \cite{Liu}, is compatible with it. Other
channels, such as $N\pi\pi$, are also open at the $\eta$ threshold and these
increase the inelasticity. We take ${\rm Im}[a\, (\eta N)] = 0.30$~fm, though
this might be an underestimate.

The energy dependence of the $\pi^{-}\, p \rightarrow \eta\, n$  amplitude
shown in fig.~3 is then enough to determine the real part of the scattering
length within the model, leading to $a\, (\eta N) = (0.55 \pm 0.20 +
i0.30)$~fm.

The sign of the real part is ambiguous and has been chosen to be attractive to
agree with Bhalerao and Liu \cite{Liu}. It should be noted that the larger
value obtained here is not due to P-wave contamination in the cross section.
This would go the wrong way but a difference could arise from different
effective range assumptions. The puzzle could perhaps be clarified by new and
very precise measurements of $\gamma p \rightarrow \eta p$ with a monochromatic
photon beam at Mainz. Preliminary data show that the cross section remains
isotropic until quite high energies \cite{Mainz}.

Using this value of $a\, (\eta N)$ in the optical potential of
eq.~(\ref{opt}), with a Gaussian nuclear density corresponding to an rms
radius of 1.9~fm, leads to a predicted scattering length of
\begin{equation}
a\, (\eta\,^{3\!}H\!e) = (-2.31 + i2.57)\ \mbox{\normalsize fm}\: .
\end{equation}

The energy dependence of the $p\, d \rightarrow \,^{3\!}H\!e\, \eta$  cross
section resulting from the simplified FSI formula of eq.~(\ref{simple})
with this scattering length is shown as the dashed line in fig.~4 and
is compared with the pioneering SPES4 data \cite{Berger} and updated
SPES2 values \cite{Garcon}. The overall normalisation (the value of
$f_{\eta}^B$) is arbitrary. Though the good numerical agreement with the data
is perhaps a little fortuitous in view of possible corrections to the
lowest order optical potential, it nevertheless indicates that the rapid fall
of the amplitude with energy is likely to be a consequence of the strong
$\eta\,^{3\!}H\!e$ final state interaction.

Once we have the potential of eq.~(\ref{opt}) then we can calculate the phase
shift at all energies, which enables us to use the more general formula of
eq.~(2) rather than the constant scattering length version of eq.~(3).
This makes little difference, as can be seen from the solid line in fig.~4.

The sign of the real part of the scattering length indicates the possible
presence of a `$\!$`bound'$\!$' $\eta$ state for a much lighter nucleus than
suggested by Haider and Liu \cite{Haider}. It can be thought of as a
displacement of the S$_{11}$ pole to below the $\eta$ threshold through the
repeated scattering of the $\eta$ on all the nucleons in $^3$He.
Other decay channels (pionic or nucleonic) are however open and the large
imaginary component in the scattering length severely limits observation of
effects from this pole.

The S-wave FSI enhancement factor of eq.~(2) is independent of the entrance
channel, though the particular nuclear reaction would influence the amount of
P and higher waves present. It should therefore be applicable also to the
$\pi^{-\,3\!}H\!e\,\rightarrow \eta\,^{3\!}H$ reaction. Unfortunately the
lowest energy for which this has been measured \cite{Peng} corresponds to
$p_{\eta} = 0.41$~fm$^{-1}$, which is just off the scale of fig.~1~!

What happens for heavier nuclei? Data exist for $d\, d \rightarrow
\,^{4\!}H\!e\,\eta$, but only away from threshold \cite{LG}. Taking an rms
radius of 1.63~fm, the $\eta\,^{4}$He potential is stronger but of shorter
range than for $\eta\,^{3}$He. The predicted scattering length
of $(-2.00 + i0.97)$~fm corresponds to a somewhat less steep energy
dependence than for $\eta\,^{3}$He production. Including effective range
effects through eq.~(2), the decrease in $|f|^2$ between $p_{\eta}$ = 0.1 and
0.4 fm$^{-1}$ is expected to be about 2.8 for $^3$He but only 1.9 for $^4$He.

Coherent $\eta$ production has been measured in
$p\,^{6\!}Li \rightarrow \eta\,^{7\!}Be^{*}$ \cite{Beppe}, though the
resolution obtained by detecting the $\eta$ through its 2$\gamma$ decay was
insufficient to isolate individual states in the $^{7}$Be nucleus. Since the
optical potential of eq.~(\ref{opt}) predicts a scattering length of
$(-2.92 + i1.21)$~fm and the typical $\eta$ centre-of-mass momentum in this
experiment was $p_{\eta}\sim 0.5\ $fm$^{-1}$, these data lie
\underline{outside} the FSI peak.

Turning now to lighter systems, Ueda suggested \cite{Ueda} that, even for a
system as diffuse as the deuteron, the S$_{11}$ pole should be moved towards
the $\eta$ threshold. Using a separable coupled channel model, with $\pi$ and
$\eta$ exchange, he predicted a scattering length $a_{\eta d}\approx
(1.8 + i2.5)$~fm.

In their analysis of old $np\rightarrow d\eta$ data, the authors of
ref.~\cite{PFW} assumed that the cross section was modulated by an {\it ad hoc}
function of the form
\begin{equation}
\label{PFW}
\frac{1}{(1+b^2p_{\eta}^2)^2}\: .
\end{equation}
In order to explain the {\it shape} of the spectrum, a large value of $b$
was required ($\approx 3$~fm), but this was not a dedicated experiment and the
available range in $p_{\eta}$ was rather small. However evidence for the size
of $b$ can be obtained in another way.

It has been shown that at 1.3 GeV the yield of $\eta$'s in $pn$
collisions is about a factor of 10 larger than in $pp$ \cite{Beppe2}.
Furthermore there has recently been an independent measurement of the
$pp \rightarrow pp\eta$ total cross section near threshold \cite{Amnon}.
Using this as normalisation, the combined experiments suggest that
$\sigma_{T}(np\rightarrow \eta\,X) \approx 0.03$~mb at $p_{\eta}\approx
0.6$~fm$^{-1}$. If this is dominated by deuteron formation, then the
combination with the threshold {\it normalisation} claimed in
ref.~\cite{PFW} implies that $b$ is at least 3~fm.

Theoretical predictions $pp \rightarrow pp\eta$ total cross section
largely follow a phase space behaviour in terms of the energy $Q_{cm}$
above threshold \cite{X}. To fit the energy dependence of the three
experimental points of ref.~\cite{Amnon}, the authors used the functional
form of eq.~(\ref{PFW}) with the {\it same} value of $b=3$~fm.

Though none of these nucleon-nucleon determinations is yet compelling, they do
seem to support Ueda's contention \cite{Ueda} that the S$_{11}$ pole is
significantly displaced already in the two-nucleon sector for both
isospin-zero {\it and} one. Unfortunately neither of these states is coupled
to the $\pi d$ system.

In summary, there is ample evidence that the low energy $\eta$ interaction
with the few-nucleon system is very strong but much more experimental work
is needed to pin it down in detail. In particular I would suggest the
measurement of the following near-threshold reactions:
\begin{itemize}
\item
$np\rightarrow d\eta$. This might be done using $pd\rightarrow d\,p_s\,\eta$
with the cooled proton beam and gas jet target at Celsius. The spectator
proton $p_s$ could then be picked up in a solid state counter.
\item
More points are needed in the $pp\rightarrow pp\eta$ excitation function.
\item
The separation of the real and imaginary parts of the scattering length
in the $\eta^3$He case could be done through measurements at Celsius at
fractions of MeV above threshold.
\item
The $\pi^{-\,3\!}H\!e\,\rightarrow \eta\,^{3\!}H$ of ref.~\cite{Peng} should
be measured closer to threshold. The results might not be quite as clear as
for the proton beams since the pion reactions could be more peripheral.
\item
$d\, d \rightarrow \,^{4\!}H\!e\,\eta$ looks like a strong possibility at
Saturne.
\item
$p\,^{3\!}H \rightarrow \,^{4\!}H\!e\,\eta$ has been foiled at LAMPF
but $n\,^{3\!}H\!e \rightarrow \,^{4\!}H\!e\,\eta$ might be easier.
\item
In the $p\,^{6\!}Li \rightarrow \eta\,^{7\!}Be^{*}$ case \cite{Beppe},
it might be advantageous to study the cross sections at say
1--2 MeV above the threshold for the excitation of a particular level.
\end{itemize}

What should theoreticians do while waiting for the results from some of
the above list?
\begin{itemize}
\item
A study of more refined models of the S$_{11}$ and other resonances which
decay into $\eta N$. One difficulty is that the S$_{11}$ is strongly coupled
to $\pi N$, $\eta N$ {\it and} $\rho N$.
\item
More refined models of the $\eta NN$ system where the $\rho$ coupling
has to be included.
\item
Better estimates of $f_{\eta}^B$ in the $p\, d\rightarrow\,^{3\!}H\!e\,\eta$
case so that one can insert the FSI's more reliably, without having to make
the constant scattering length approximation in the Watson FSI factor.
Progress has already been
made on semi-empirical three-nucleon models \cite{Kilian,CW}.
\end{itemize}

However in the end theoreticians do what they want to and experimentalists do
what they can get money for!

Discussions with  R.Kessler and A.Moalem on the results of ref.~\cite{Garcon}
and ref.~\cite{Amnon} respectively were much appreciated. \\

\vspace{1cm}
\noindent
{\large{\bf Figure Captions}}\\[1ex]
\begin{quote}
Fig.~1: Fits to the $p\, d \rightarrow \,^{3\!}H\!e\, \eta$  data of
ref.~\cite{Garcon} with complex scattering lengths a = (3.1 + i2.5)~fm (solid
curve) and (5.0 + i0.2)~fm (dashed curve). The Laget and Lecolley predictions
(LL) \cite{LL} fall off much too slowly with $p_{\eta}$.\\[3ex]
Fig.~2: The counting rate (solid curve) at the lowest energy point of
ref.~\cite{Garcon} as a function of $\Delta T_p$, the energy above threshold
in the $p\, d \rightarrow \,^{3\!}H\!e\, \eta$  reaction. This is a product of
the cross section and experimental acceptance.\\[3ex]
Fig.~3: The square of the $\pi^{-}\, p \rightarrow \eta\, n$  amplitude
extracted from the total cross section data of ref.~\cite{Binnie}, as a
function of $p_{\eta}$. The solid line is a fit using eq.~(4) with the
imaginary part of the $\eta N$ scattering length constrained by unitarity,
leading to (0.55 + i0.30)~fm. The dashed line is the best fit with
${\rm Im}[a\, (\eta N)] = 0$.\\[3ex]
Fig.~4: Predictions for the square of the $p\, d \rightarrow \,^{3\!}H\!e\,
\eta$ amplitude using the optical potential and eq.~(3) (solid curve) and its
approximation by eq.~(4) (dashed). The pioneering data of SPES4 \cite{Berger}
are shown as crosses and those of SPES2 \cite{Garcon} as circles.
\end{quote}
\end{document}